\documentclass[showpacs,amssymb,preprintnumbers,
nofootinbib,superscriptaddress]{revtex4}

\usepackage{amsmath}
\usepackage{graphicx}
\usepackage{latexsym}
\usepackage{amsfonts}
\usepackage{url,hyperref}
\usepackage{bm}

\begin{document}

\begin{figure}[t]
\vspace{-1.4cm}
\hspace{-16.25cm}
\scalebox{0.09}{\includegraphics{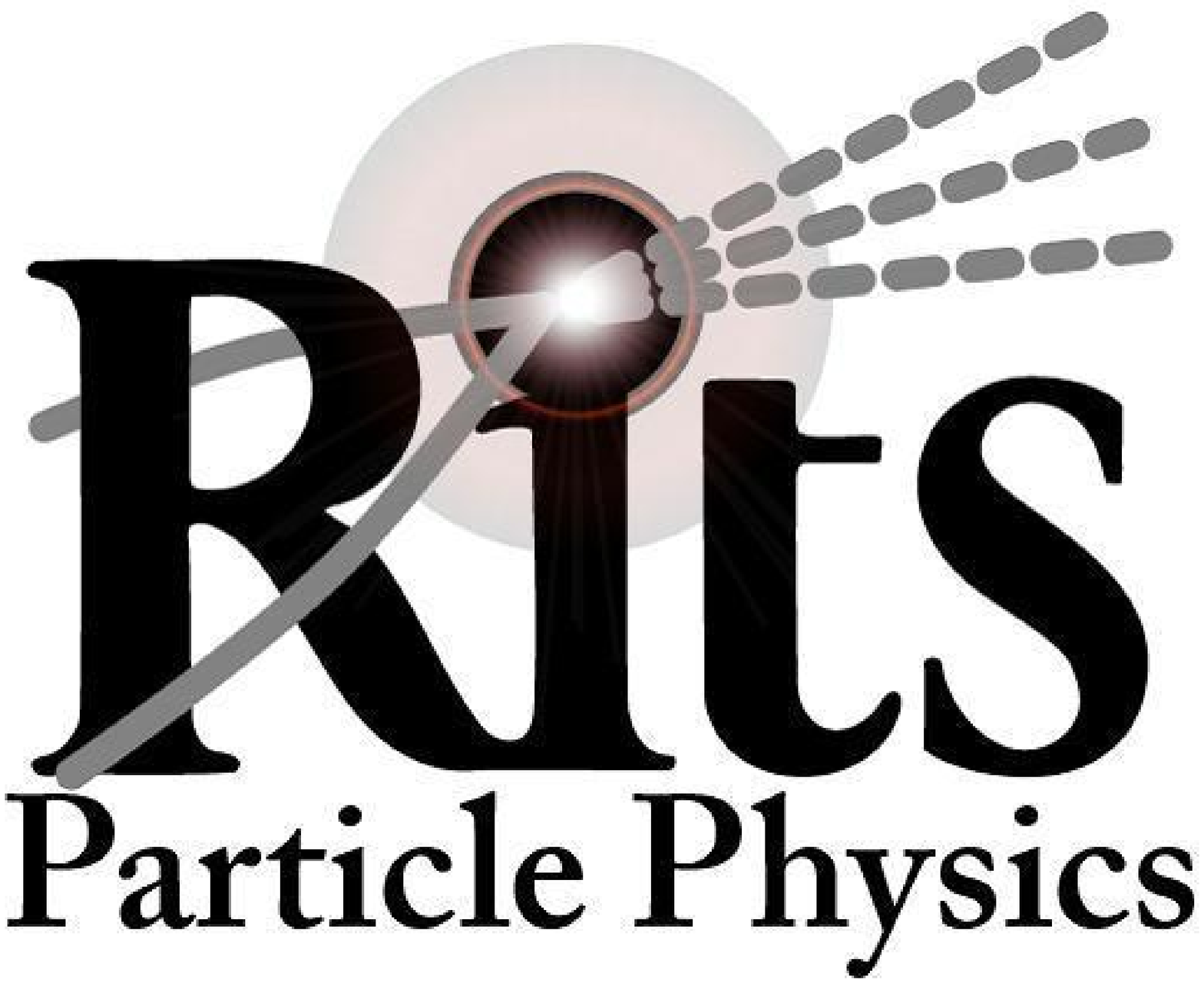}}
\end{figure}

\newcommand{\vp}{\varphi}
\newcommand{\nn}{\nonumber\\}
\newcommand{\beq}{\begin{equation}}
\newcommand{\eeq}{\end{equation}}
\newcommand{\bed}{\begin{displaymath}}
\newcommand{\eed}{\end{displaymath}}
\def\bea{\begin{eqnarray}}
\def\eea{\end{eqnarray}}
\newcommand{\veps}{\varepsilon}

\title{Split fermion quasi-normal modes}
\author{H.~T.~Cho}
\email[Email: ]{htcho``at"mail.tku.edu.tw} \affiliation{Department
of Physics, Tamkang University, Tamsui, Taipei, Taiwan, Republic
of China}
\author{A.~S.~Cornell}
\email[Email: ]{cornell``at"ipnl.in2p3.fr}
\affiliation{Universit\'e de Lyon, Villeurbanne, F-69622, France;
Universit\'e Lyon 1, Institut de Physique Nucl\'eaire de Lyon}
\author{Jason~Doukas}
\email[Email: ]{j.doukas``at"physics.unimelb.edu.au}
\affiliation{School of Physics, University of Melbourne,
Parkville, Victoria 3010, Australia.}
\author{Wade~Naylor}
\email[Email: ]{naylor``at"se.ritsumei.ac.jp}
\affiliation{Department of Physics, Ritsumeikan University,
Kusatsu, Shiga 525-8577, Japan}

\begin{abstract}
In this paper we use the conformal properties of the spinor field
to show how we can obtain the fermion quasi-normal modes for a
higher dimensional Schwarzschild black hole. These modes are of
interest in so called split fermion models, where quarks and
leptons are required to exist on different branes in order to keep
the proton stable. As has been previously shown, for brane
localized fields, the larger the number of dimensions the faster
the black hole damping rate. Moreover, we also present the
analytic forms of the quasi-normal frequencies in both the large
angular momentum and the large mode number limits.
\end{abstract}

\preprint{LYCEN 2006-24}
\preprint{RITS-PP-011}
\date{\today}
\maketitle

\section{Introduction}

\par With the advent of theories postulating the existence of additional dimensions,
there has been much discussion in the literature related to the
quasi-normal modes (QNMs) of black holes (BHs), for example, in
the context of the QNMs of higher dimensional BHs see reference
\cite{HiQNM}. By QNMs we refer to the complex frequency modes of
oscillation which arise from perturbations of the BH, where the
real part represents the actual frequency of the oscillation and
the imaginary part represents the damping due to the emission of
gravitational waves.

\par Recent investigations of large extra-dimensional scenarios \cite{ADD}, where the hierarchy problem can be shifted into a problem of the scale of the extra-dimensions, has led to the somewhat striking prediction that BHs may be observed at particle accelerators such as the LHC \cite{BHacc}, for an interesting treatment of mini BHs
within an effective field theory framework see \cite{BH4}.
However, one poignant problem is that in order to suppress a rapid
proton decay we need to physically split the quarks and leptons.
Such models are generically called split fermions models, see, for
example, reference \cite{Split}. In supersymmetric versions of
this idea the localizing scalars and bulk gauge fields will have
fermionic bulk superpartners.  In this respect it is important to
consider the properties of bulk fermions. Note that up to now only
brane localized QNMs have been calculated \cite{BraneQNMs} (where
other BH effects have been considered in reference
\cite{Stojkovic}).

\par That said, the motivations for studying the fermion QNMs from BHs
in this paper are two-fold; the first of these being from the
theoretical point of view, where the lack of any work done in
greater than four dimensions \cite{Cho2} with Dirac fields is an
omission in the literature. Our calculations serve to fill this
gap. Secondly, having a complete catalog of all QNMs would be a
necessary precursor to eventually studying the emission rates of
collider produced, or TeV scale, BHs. In a forthcoming work we
shall present details of BH absorption cross-sections for bulk
Schwarzschild fermions in $d$-dimensions \cite{Forth}. See also
reference \cite{Dai} for tense branes in six-dimensions, where
bulk fermions on the tense brane background defined there have yet
to be obtained.

\par As such, this paper shall be structured as follows: In the next section we shall discuss how a conformal transformation of the metric allows for a convenient separation of the Dirac equation into a time-radial part and a $(d-2)$-sphere, in $d$-dimensions. Such a method has already been applied in reference \cite{Das} to the case of low energy $s$-wave absorption cross-sections (which the authors are currently generalizing to higher energies and quantum numbers). However, this has not been used yet in the context of BH QNMs. After this we shall present our results for the QNMs, along with the analytical forms of the frequencies in both the large angular momentum as well as the large mode number limits. Finally, in the last section, we shall make some concluding statements.

\section{Spinor radial wave equation}

\par We shall begin our analysis by supposing a background metric which is $d$-dimensional and spherically symmetric, as given by:
\begin{equation}
ds^{2}=-f(r)dt^{2}+h(r)dr^{2}+r^{2}d\Omega^{2}_{d-2} ,
\end{equation}
where $d\Omega^{2}_{d-2}$ denotes the metric for the $(d-2)$-dimensional sphere.

\par Under a conformal transformation \cite{Das,Gibbons}:
\begin{eqnarray}
g_{\mu\nu} & \rightarrow & \overline{g}_{\mu\nu}=\Omega^{2}g_{\mu\nu} , \\
\psi & \rightarrow & \overline{\psi}=\Omega^{-(d-1)/2}\psi , \\
\gamma^{\mu}\nabla_{\mu}\psi & \rightarrow & \Omega^{(d+1)/2} \overline{\gamma}^{\mu}\overline{\nabla}_{\mu}\overline{\psi} ,
\end{eqnarray}
where we shall take $\Omega=1/r$, the metric becomes:
\beq
d\overline{s}^{2} = -\frac{f}{r^{2}}dt^{2} + \frac{h}{r^{2}}dr^{2} + d\Omega^{2}_{d-2} , \qquad\qquad \mathrm{where} \qquad \overline{\psi}=r^{(d-1)/2}\psi . \\
\eeq
Since the $t-r$ part and the $(d-2)$-sphere part of the metric are completely separated, one can write the Dirac equation in the form:
\begin{eqnarray}
& \overline{\gamma}^{\mu}\overline{\nabla}_{\mu}\overline{\psi}=0
& \nonumber\\ & \Rightarrow \left[ \left(
\overline{\gamma}^{t}\overline{\nabla}_{t} +
\overline{\gamma}^{r}\overline{\nabla}_{r} \right) \otimes 1
\right] \overline{\psi} + \left[ \overline{\gamma}^{5} \otimes
\left( \overline{\gamma}^{a}
\overline{\nabla}_{a}\right)_{S_{d-2}} \right] \overline{\psi} = 0
, &
\end{eqnarray}
where $(\overline{\gamma}^{5})^{2}=1$. Note that from this point on we shall change our notation by omitting the bars.

\par We shall now let $\chi_{l}^{(\pm)}$ be the eigenspinors for the $(d-2)$-sphere \cite{Camporesi}, that is:
\begin{equation}
\left( \gamma^{a}\nabla_{a} \right)_{S_{d-2}}\chi_{l}^{(\pm)} = \pm i \left( l + \frac{d-2}{2}\right) \chi_{l}^{(\pm)} ,
\end{equation}
where $l = 0, 1, 2, \dots$. Since the eigenspinors are orthogonal, we can expand $\psi$ as:
\begin{equation}
\psi = \sum_{l} \left( \phi_{l}^{(+)} \chi_{l}^{(+)} + \phi_{l}^{(-)} \chi_{l}^{(-)} \right) .
\end{equation}
The Dirac equation can thus be written in the form:
\begin{equation}\label{eqn:2Ddirac}
\left \{ \gamma^{t} \nabla_{t} + \gamma^{r} \nabla_{r} + \gamma^{5} \left[ \pm i \left( l + \frac{d-2}{2} \right) \right] \right\} \phi_{l}^{(\pm)} = 0 ,
\end{equation}
which is just a $2$-dimensional Dirac equation with a $\gamma^{5}$ interaction.

\par To solve this equation we make the explicit choice of the Dirac matrices:
\begin{equation}
\gamma^{t} = \frac{r}{\sqrt{f}}(-i\sigma^{3})\ \ \ ,\ \ \
\gamma^{r}=\frac{r}{\sqrt{h}}\sigma^{2} ,
\end{equation}
where the $\sigma^{i}$ are the Pauli matrices:
\begin{equation}
\sigma^{1}=\left(
\begin{array}{cc}
0 & 1 \\ 1 & 0
\end{array}
\right)\ \ \ ,\ \ \ \sigma^{2}=\left(
\begin{array}{cc}
0 & -i \\ i & 0
\end{array}
\right)\ \ \ ,\ \ \ \sigma^{3}=\left(
\begin{array}{cc}
1 & 0 \\ 0 & -1
\end{array}
\right) .
\end{equation}
Also,
\begin{equation}
\gamma^{5} = (-i\sigma^{3})(\sigma^{2}) = - \sigma^{1} .
\end{equation}
The spin connections are then found to be:
\beq
\Gamma_{t} = \sigma^{1} \left( \frac{r^{2}}{4\sqrt{fh}} \right) \frac{d}{dr} \left( \frac{f}{r^{2}} \right) , \qquad\qquad \Gamma_{r} = 0 .
\eeq

\par From this point on we shall work with the $+$ sign solution, where the $-$ sign case would work in the same way. The Dirac equation can then be written explicitly as:
\begin{eqnarray}
& \displaystyle \left\{ \frac{r}{\sqrt{f}}(-i\sigma^{3}) \left[
\frac{\partial}{\partial t} + \sigma^{1} \left(
\frac{r^{2}}{4\sqrt{fh}} \right) \frac{d}{dr} \left(
\frac{f}{r^{2}} \right) \right] + \frac{r}{\sqrt{h}} \sigma^{2}
\frac{\partial}{\partial r} + (-\sigma^{1})(i) \left( l +
\frac{d-2}{2} \right) \right\} \phi_{l}^{(+)} = 0  & \nonumber\\ &
\displaystyle \Rightarrow \sigma^{2} \left( \frac{r}{\sqrt{h}}
\right) \left[ \frac{\partial}{\partial r} + \frac{r}{2\sqrt{f}}
\frac{d}{dr} \left( \frac{\sqrt{f}}{r} \right) \right]
\phi_{l}^{(+)} - i \sigma^{1} \left( n + \frac{d-2}{2} \right)
\phi_{l}^{(+)} = i \sigma^{3} \left( \frac{r}{\sqrt{f}} \right)
\frac{\partial \phi_{l}^{(+)}}{\partial t} . &
\end{eqnarray}
We shall now determine solutions of the form:
\begin{equation}
\phi_{l}^{(+)} = \left( \frac{\sqrt{f}}{r} \right)^{-1/2} e^{-iEt} \left(
\begin{array}{c}
iG(r) \\ F(r)
\end{array}
\right) ,
\end{equation}
where $E$ is the energy. The Dirac equation can then be simplified to:
\begin{equation}
\sigma^{2} \left( \frac{r}{\sqrt{h}} \right) \left(
\begin{array}{c}
i\frac{dG}{dr} \\ \frac{dF}{dr}
\end{array}
\right) -i \sigma^{1} \left( l + \frac{d-2}{2} \right) \left(
\begin{array}{c}
iG \\ F
\end{array}
\right) = i \sigma^{3}E \left( \frac{r}{\sqrt{f}} \right) \left(
\begin{array}{c}
iG \\ F
\end{array}
\right) ,
\end{equation}
or
\begin{eqnarray}
\sqrt{\frac{f}{h}} \frac{dG}{dr} - \frac{\sqrt{f}}{r} \left( l +
\frac{d-2}{2} \right) G & = & EF \\ \sqrt{\frac{f}{h}}
\frac{dF}{dr} + \frac{\sqrt{f}}{r} \left( l + \frac{d-2}{2}
\right) F & = & - EG .
\end{eqnarray}
It will be convenient to define the tortoise coordinate, $r_*$,
and the function $W$ as: \beq
\sqrt{\frac{f}{h}}\frac{d}{dr}\equiv\frac{d}{dr_{*}} ,
\qquad\quad\qquad W=\frac{\sqrt{f}}{r}\left(l+\frac{d-2}{2}\right)
. \eeq In which case, our equations can be expressed as: \beq
\left(\frac{d}{dr_{*}}-W\right)G=EF , \qquad\qquad
\left(\frac{d}{dr_{*}}+W\right)F=-EG . \eeq The equations can then
be separated to give: \beq
\left(-\frac{d}{dr_{*}^{2}}+V_{1}\right)G=E^{2}G \qquad
\mathrm{and} \qquad
\left(-\frac{d}{dr_{*}^{2}}+V_{2}\right)F=E^{2}F
,\label{gequation}
 \eeq where:
\begin{equation}
V_{1,2}=\pm\frac{dW}{dr_{*}}+W^{2} .
\end{equation}
Since $V_{1}$ and $V_{2}$ are supersymmetric to each other, $F$ and $G$ will have the same spectra, both for scattering and quasi-normal. Incidentally, for $\phi_{n}^{(-)}$, we have
these two potentials again.

\par Refining our study now to a $d$-dimensional Schwarzschild BH, where $f$ becomes:
\begin{equation}
f(r)=h^{-1}(r)=1-\left(\frac{r_{H}}{r}\right)^{d-3} ,
\end{equation}
and where the horizon is at $r=r_{H}$ with:
\begin{equation}
r_{H}^{d-3}=\frac{8\pi M\Gamma((d-1)/2)}{\pi^{(d-1)/2}(d-2)} .
\end{equation}
In this case the potential $V_{1}$ can be expressed as:
\begin{eqnarray}
V_{1}(r)
&=&f\frac{d}{dr}\left[\sqrt{f}\left(\frac{l+\frac{d-2}{2}}{r}\right)\right]
+f\left(\frac{l+\frac{d-2}{2}}{r}\right)^{2}\nonumber\\
&=&\left[1-\left(\frac{r_{H}}{r}\right)^{d-3}\right]
\frac{d}{dr}\left[\sqrt{1-\left(\frac{r_{H}}{r}\right)^{d-3}}
\left(\frac{l+\frac{d-2}{2}}{r}\right)\right]
+\left[1-\left(\frac{r_{H}}{r}\right)^{d-3}\right]
\left(\frac{l+\frac{d-2}{2}}{r}\right)^{2} .
\end{eqnarray}

Next, we shall define, for notational convenience:
\begin{eqnarray}
\kappa&\equiv& l+\frac{d-2}{2} , \\
\Delta&\equiv&r^{d-3}(r^{d-3}-r_{H}^{d-3}) ,
\end{eqnarray}
with $\kappa = \frac{d}{2}-1, \frac{d}{2}, \frac{d}{2}+1, \dots$. This allows the above potential, $V_{1}$, to be simplified to:
\begin{equation}
V_{1}=\frac{\kappa\Delta^{1/2}}{r^{2(d-2)}}\left[\kappa\Delta^{1/2}
-r^{d-3}+\left(\frac{d-1}{2}\right)r_{H}^{d-3}\right] .
\label{vone}
\end{equation}
For the case $d=4$, this then becomes:
\begin{equation}
V_{1}=\frac{\kappa\Delta^{1/2}}{r^{4}}\left(\kappa\Delta^{1/2}-r+3M\right) ,
\end{equation}
with $\Delta=r(r-2M)$, which is just the radial equation for the
Dirac equation in the 4-dimensional Schwarzschild BH \cite{Cho}.

\section{QNMs Using The Iyer and Will Method}


\begin{table}
\caption{Massless bulk Dirac QNM frequencies (${\rm Re}(E)>0$) for a higher dimensional Schwarzschild BH with $l\geq n\geq 0$, with $d=5,6,7,8,9$ and $10$ dimensions. Given the
accuracy of the 3rd order WKB method, results are presented up to three significant figures only.}
\label{Dirac}
\begin{ruledtabular}
\begin{tabular}{cccc}
$(l,n)$ \quad Odd $d$ &  $d=5$  & $d=7$ & $d=9$ \\
\hline
$l$=0, n=0 & 0.725 - 0.396 i & 1.79 - 0.809 i & 2.66 - 0.999 i \\
$l$=1, n=0 & 1.32 - 0.384 i & 2.73 - 0.807 i & 3.73 - 1.03 i \\
$l$=1, n=1 & 1.15 - 1.22 i & 2.05 - 2.68 i & 2.30 - 3.57 i \\
$l$=2, n=0 & 1.88 - 0.384 i & 3.61 - 0.817 i & 4.71 - 1.06 i \\
$l$=2, n=1 & 1.75 - 1.18 i & 3.11 - 2.56 i & 3.65 - 3.35 i \\
$l$=2, n=2 & 1.56 - 2.03 i & 2.27 - 4.53 i & 1.80 - 6.23 i \\
$l$=3, n=0 & 2.43 - 0.384 i & 4.46 - 0.821 i  & 5.64 - 1.08 i \\
$l$=3, n=1 & 2.33 - 1.17 i & 4.08 - 2.52 i & 4.85 - 3.29 i \\
$l$=3, n=2 & 2.17- 1.99 i & 3.38 - 4.36 i & 3.30 - 5.84 i \\
$l$=3, n=3 & 1.96 - 2.84 i & 2.46 - 6.36 i  & 1.30 - 8.87 i \\
\hline
$(l,n)$ \quad Even $d$ & $d=6$ & $d=8$ & $d=10$ \\
\hline
$l$=0, n=0 & 1.28 - 0.639 i  & 2.24 - 0.924 i & 3.05 - 1.05 i \\
$l$=1, n=0 & 2.10 - 0.623 i & 3.27 - 0.936 i & 4.14 - 1.10 i \\
$l$=1, n=1 & 1.71 - 2.04 i & 2.24 - 3.18 i & 2.26 - 3.87 i \\
$l$=2, n=0 &  2.87- 0.631 i  &  4.21 - 0.956 i & 5.14 - 1.14 i \\
$l$=2, n=1 & 2.11 - 3.42 i & 3.45 - 3.01 i & 3.75 - 3.59 i \\
$l$=2, n=2 & 2.27 - 4.53 i & 2.15 - 5.45 i & 1.26 - 6.95 i \\
$l$=3, n=0 & 3.61 - 0.632 i & 5.11 - 0.964 i & 6.08 - 1.16 i \\
$l$=3, n=1 & 3.39 - 1.93 i  & 4.54 - 2.96 i & 5.06 - 3.54 i \\
$l$=3, n=2 &3.00 - 3.32 i  & 3.46 - 5.18 i & 2.95 - 6.340 i \\
$l$=3, n=3 & 2.49 - 4.78 i & 2.04 - 7.69 i & 0.327 - 10.0 i
\end{tabular}
\end{ruledtabular}
\end{table}
\par To evaluate the QNM frequencies we adopt the WKB approximation developed by Iyer and Will \cite{IW}, also see references therein. Note that this analytic method has been used extensively in various BH cases \cite{Iyer}, where comparisons with other numerical results have been found to be accurate up to around 1\% for both the real and the imaginary parts of the frequencies for low-lying modes with $n<l$ (where $n$ is the mode number and $l$ is the spinor angular momentum quantum number). Furthermore, we have also included the $n=l$ modes in our results, but the inclusion of these modes does depend on the number of the dimensions $d$. The formula for the complex quasi-normal mode frequencies $E$ in the WKB approximation, carried to third order beyond the eikonal approximation, is given by \cite{IW}:
\begin{equation}
E^{2}=[V_{0}+(-2V^{''}_{0})^{1/2}\Lambda]
-i(n+\frac{1}{2})(-2V^{''}_{0})^{1/2}(1+\Omega) ,
\label{WKBeq}
\end{equation}
where we denote $V_0$ as the maximum of $V_1$ and
\begin{eqnarray}
\Lambda&=&\frac{1}{(-2V^{''}_{0})^{1/2}}
\left\{\frac{1}{8}\left(\frac{V_{0}^{(4)}}{V_{0}^{''}}\right)
\left(\frac{1}{4}+\alpha^{2}\right)- \frac{1}{288}
\left(\frac{V_{0}^{'''}}{V_{0}^{''}}\right)^{2}(7+60\alpha^{2})\right\},\\
\Omega&=&\frac{1}{(-2V_{0}^{''})}\left\{\frac{5}{6912}
\left(\frac{V_{0}^{'''}}{V_{0}^{''}}\right)^{4}(77+188\alpha^{2})
-\frac{1}{384}\left(\frac{{V_{0}^{'''}}^{2}V_{0}^{(4)}}{{V_{0}^{''}}^{3}}\right)
(51+100\alpha^{2})\right.\nonumber\\ &&\ \ \
+\frac{1}{2304}\left(\frac{V_{0}^{(4)}}{V_{0}^{''}}\right)^{2}(67+68\alpha^{2})
+\frac{1}{288}\left(\frac{V_{0}^{'''}V_{0}^{(5)}}{{V_{0}^{''}}^{2}}\right)(19+28\alpha^{2})
\left.
-\frac{1}{288}\left(\frac{V_{0}^{(6)}}{V_{0}^{''}}\right)(5+4\alpha^{2})\right\} .
\end{eqnarray}
Here
\begin{eqnarray}\label{eqn:alpha}
\alpha&=&n+\frac{1}{2},\ n=\left\{
\begin{array}{l}
0,1,2,\cdots,\ {\rm Re}(E)>0\\ -1,-2,-3,\cdots,\ {\rm Re}(E)<0
\end{array}
\right.
\qquad \mathrm{and} \qquad V_{0}^{(n)}=\left.\frac{d^{n}V}{dr_{\ast}^{n}}\right|_{r_{\ast}=r_{\ast}(r_{max})} .
\end{eqnarray}

\begin{figure}[t]
\begin{center}
\scalebox{0.75}{\includegraphics{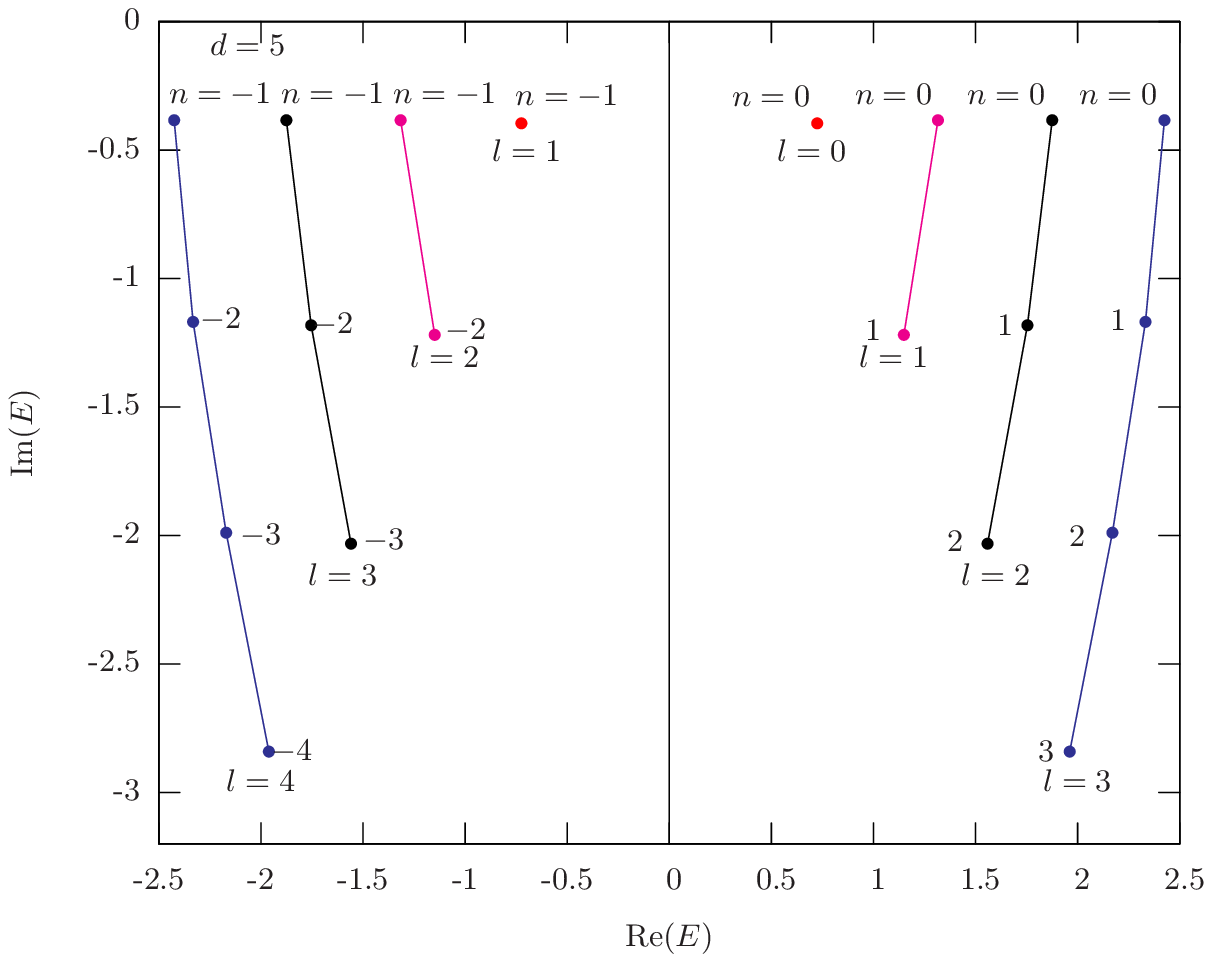}}
\vspace{0.25cm}
\scalebox{0.75}{\includegraphics{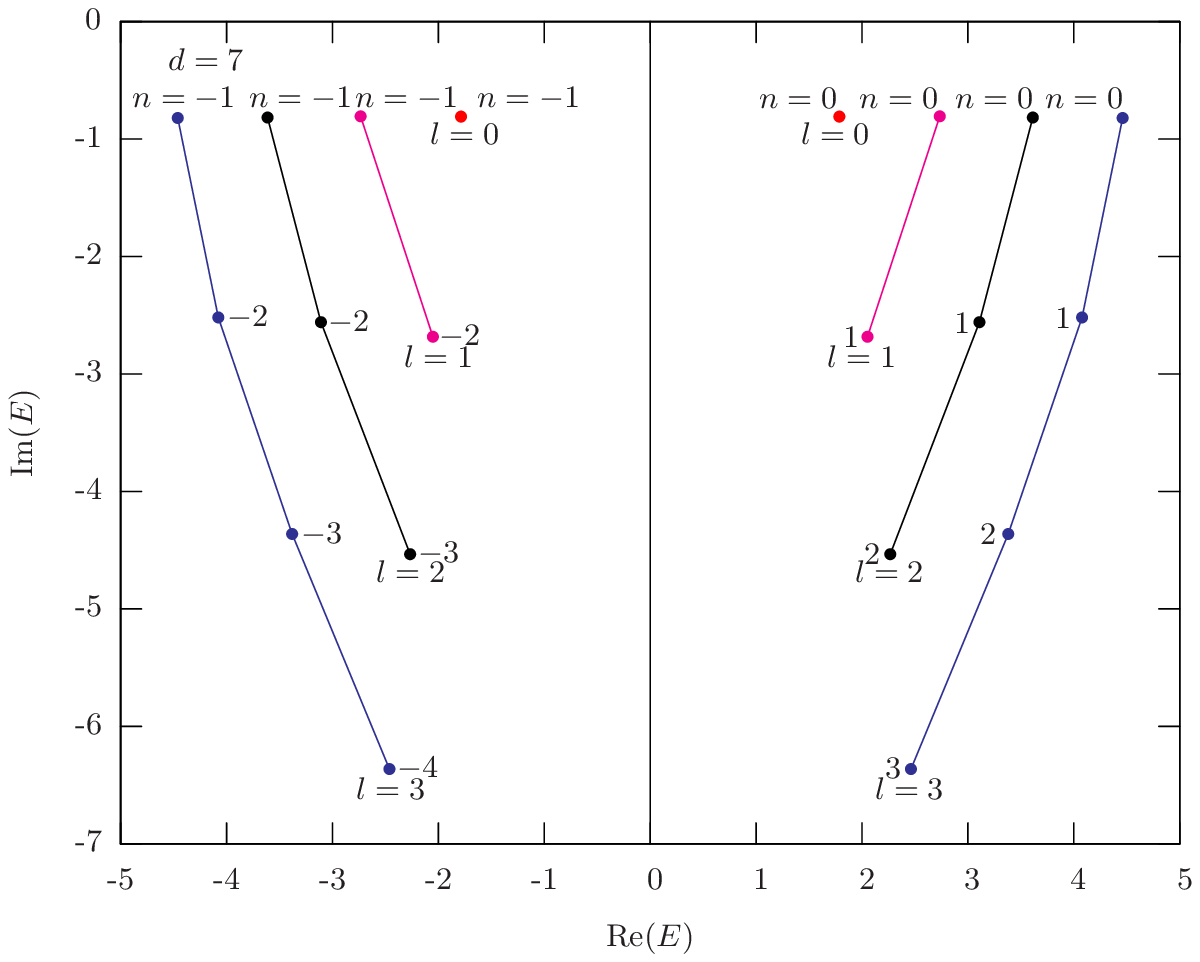}}
\scalebox{0.75}{\includegraphics{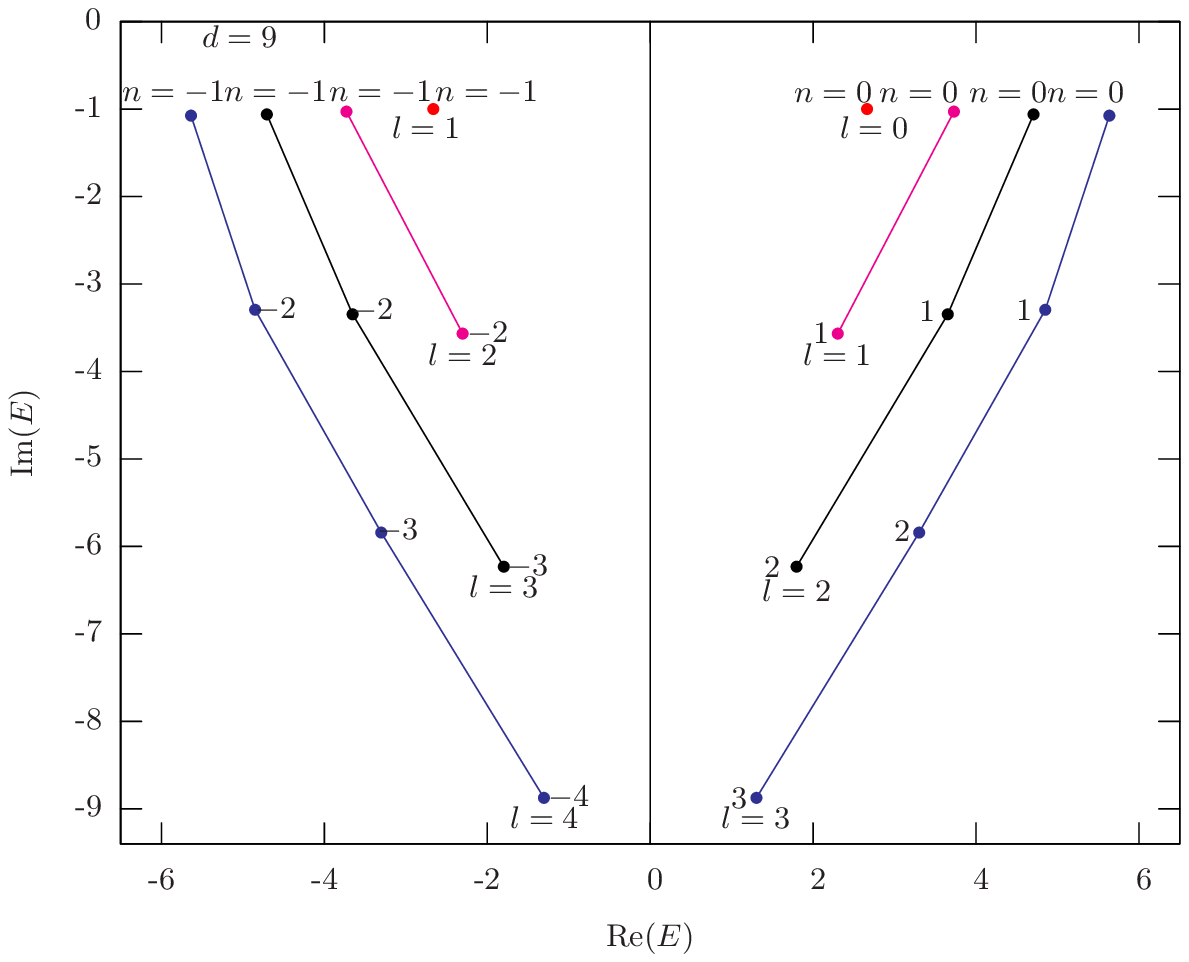}}
\end{center}
\caption{Lines of constant $l$ for massless bulk Dirac QNM frequencies for a Schwarzschild BH in odd $d$-dimensions.}
\label{QNModd}
\end{figure}

\begin{figure}[t]
\begin{center}
\scalebox{0.75}{\includegraphics{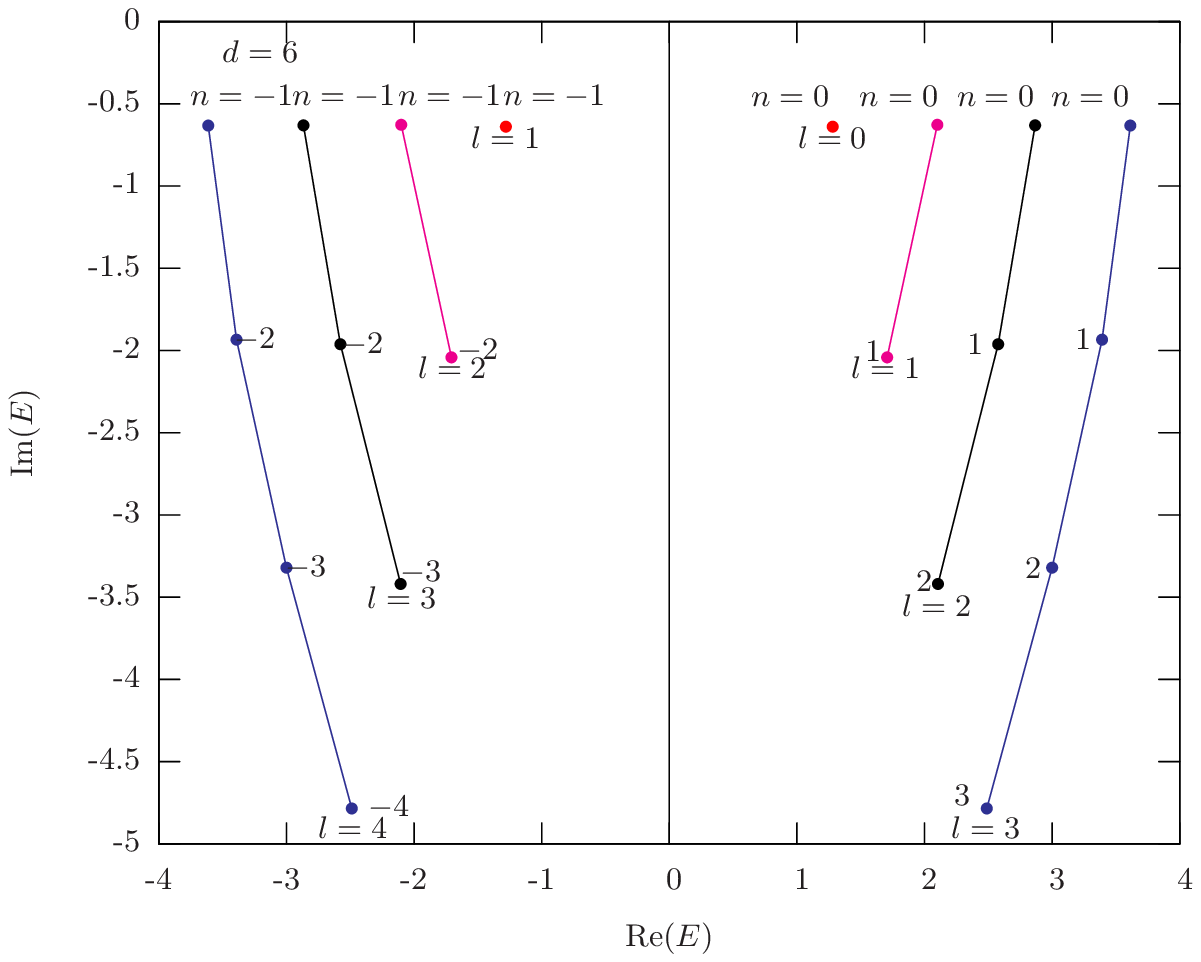}}
\vspace{0.25cm}
\scalebox{0.75}{\includegraphics{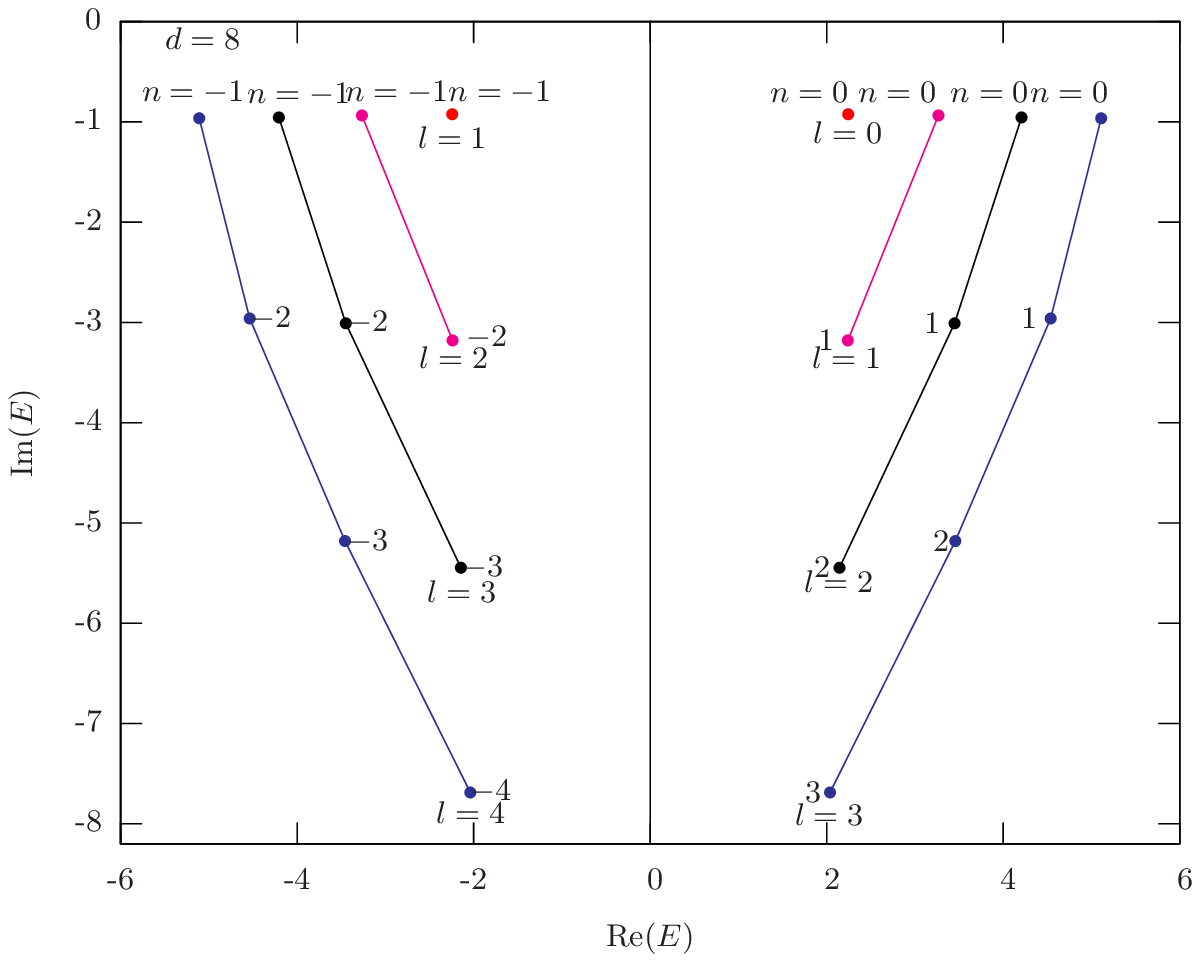}}
\scalebox{0.75}{\includegraphics{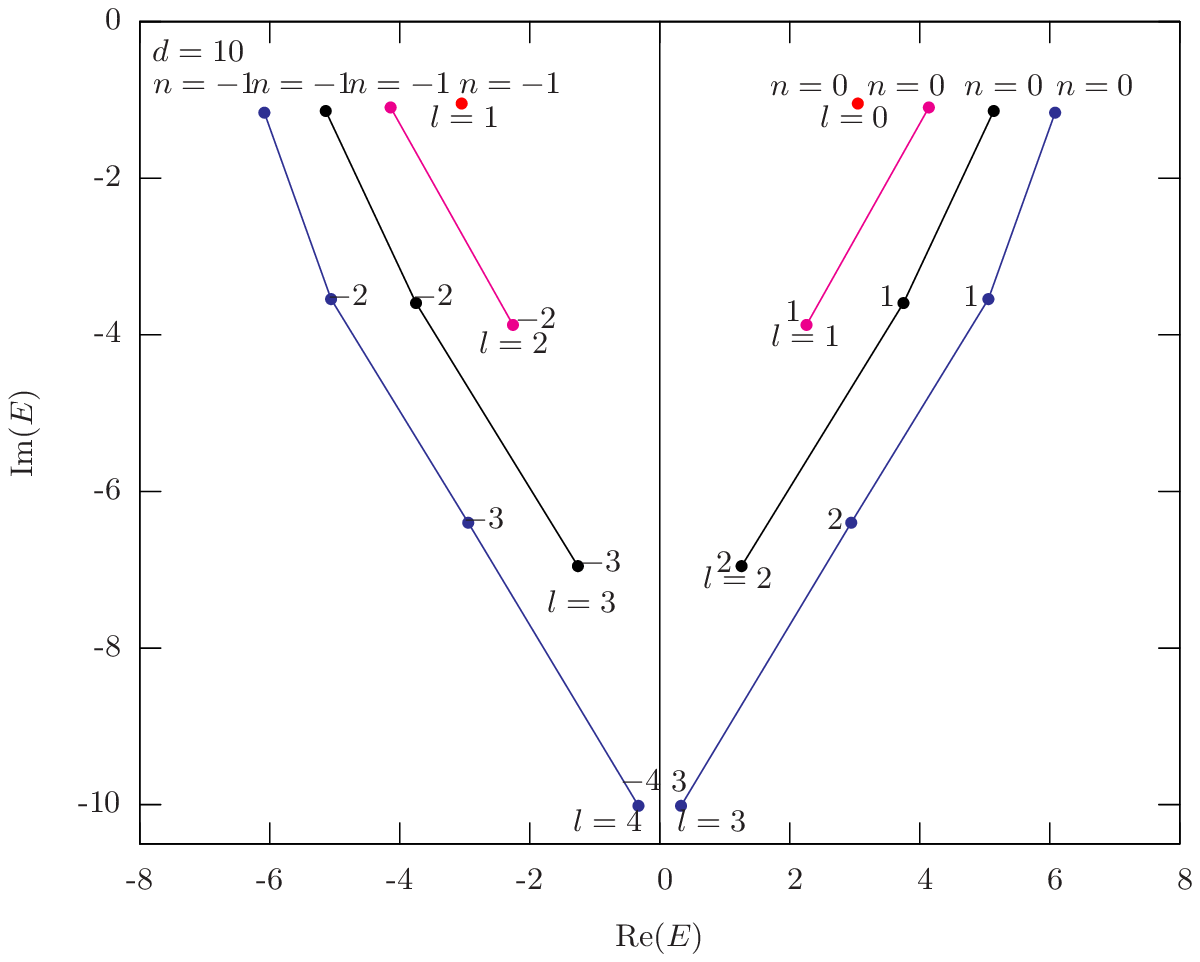}}
\end{center}
\caption{Lines of constant $l$ for massless bulk Dirac QNM frequencies for a Schwarzschild BH in even $d$-dimensions.}
\label{QNMeven}
\end{figure}

\begin{figure}[!]
\includegraphics{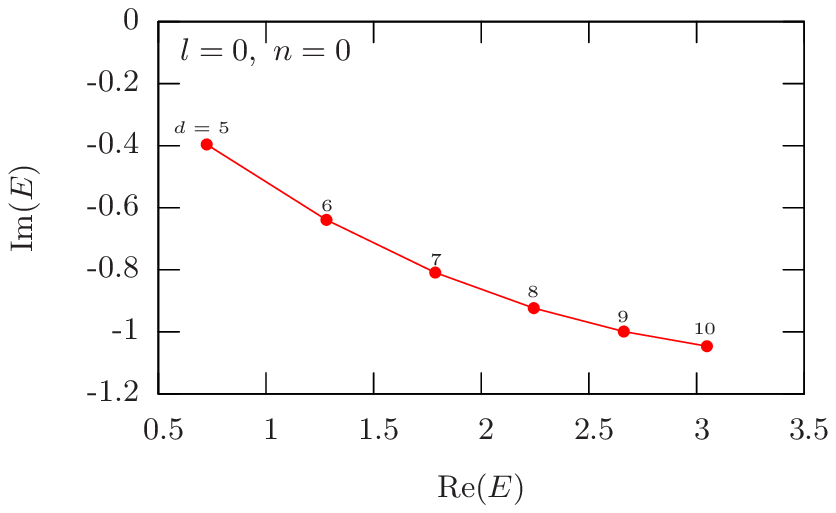}
\caption{Variation of the QNM frequency with dimension for
$l=n=0$. Both the frequency of oscillation and the damping rate
increase with the number of dimensions.}
\label{QNMwithdim}
\end{figure}


\par It is worth mentioning that in the spin-1/2 case it does not seem possible to solve for $r_{max}$ for an arbitrary value of $d$, that is, to find an analytic solution for the roots (unlike the case for fields of other spins \cite{Iyer}). In the case of a bulk spin-0 field and the graviton tensor perturbations on a $d$-dimensional Schwarzschild background a similar analytic expression can be found for $r_{max}$ in $d$ dimensions, see reference \cite{Cornell}. Thus, we must find this maximum numerically using a root finding algorithm. Note that for a given $d$ we can solve for the roots analytically using a symbolic computer program, although this is not essential, it does improve the performance of our code.

\section{Large Angular Momentum}

\par If we now focus on the large angular momentum limit ($\kappa\to\infty$) we can easily extract an analytic expression for the QNMs to first order:
\beq E^{2}\approx V_{0} -i(n+\frac{1}{2})(-2V^{''}_{0})^{1/2} +
\dots , \label{largeK} \eeq where $V_0$ is the maximum of the
potential $V_1$, see equation (\ref{vone}). In this limit the
potential now takes the form: \beq V_{1} \Big|_{\kappa\to\infty}
\approx {\kappa^2 (r^{d-3}-r^{d-3}_h) \over r^{d-1}} . \eeq The
location of the maximum of the potential is at \beq r_{max}
\Big|_{\kappa\to\infty} \approx \left(d-1\over 2\right)^{1\over
d-3}\, r_H . \eeq The maximum of the potential in such a limit is
then found to be:
\begin{equation}
V_{0}\Big|_{\kappa\to\infty}\approx\frac{\kappa^{2}2^{\frac{2}{d-3}}(d-3)}
{(d-1)^{\frac{d-1}{d-3}}r_{H}^{2}}.
\end{equation}
In this case we find from the 1st order WKB approximation that:
\beq E\Big|_{\kappa\to\infty} \approx
{2^{1\over{d-3}}\sqrt{d-3}\over (d-1)^{d-1\over 2(d-3)}r_H}
\left[\kappa  - i \big(n+\frac 1 2\big)\sqrt{d-3} \right] . \eeq
This result agrees with the standard result in four dimensions,
$d=4$, for example see reference \cite{Iyer}, and is similar to
the spin-0 result given in reference \cite{Konoplya}. These
limiting values also appear to agree well with the plots made in
Figure~\ref{QNModd} and \ref{QNMeven}. We have also plotted
the QNM $n=l=0$ dependence on dimension, $d$, in Figure~\ref{QNMwithdim}.

\section{Asymptotic quasinormal frequency}

\par Finally, we can also calculate the quasinormal frequency in the limit of large mode number, $n$. In this case, as we can see from the results in the previous sections, that as $n \rightarrow \infty$ the imaginary part of $E$ tends to negative infinity, ${\rm Im} E \rightarrow - \infty$. Hence, we are really looking at the large $|E|$ limit here. Using the method by Andersson and Howls \cite{Andersson}, who have combined the WKB formalism with the monodromy method of Motl and Neitzke \cite{Motl}, we make our evaluations in this limit. Note that in reference \cite{Cho2} this method has been used to obtain the asymptotic quasinormal frequency for the four-dimensional Dirac field. Since we are following the same procedure as in references \cite{Cho2,Andersson}, we shall only show the essential steps, where one can consult references \cite{Cho2,Andersson} for details. To start we return to equation~(\ref{gequation}):
\begin{equation}
\left(-\frac{d^{2}}{dr_{*}^{2}}+V_{1}\right)G=E^{2}G \Rightarrow
\frac{d^{2}G}{dr_{*}^{2}}+\left(E^{2}-V_{1}\right)G = 0 . \label{gequation2}
\end{equation}
Defining a new function,
\begin{equation}
Z(r)=\frac{\Delta^{1/2}}{r^{d-3}}G(r) ,
\end{equation}
equation~(\ref{gequation2}) can be rewritten as:
\begin{equation}
\frac{d^{2}Z}{dr^{2}}+R(r)Z=0 ,
\end{equation}
with
\begin{equation}
R(r)=\frac{r^{4(d-3)}}{\Delta^{2}}\left[E^{2}-V_{1}+\frac{(d-3)(d-2)r_{h}^{d-3}}
{2r^{d-1}}-\frac{(d-3)(d-1)r_{h}^{2(d-3)}}{4r^{2(d-2)}}\right] .
\end{equation}
The WKB solutions to this equation are:
\begin{equation}
f_{1,2}^{(t)}(r)=\frac{1}{\sqrt{Q(r)}}e^{\displaystyle \pm i\int_{t}^{r}d\xi\ \!
Q(\xi)} ,
\end{equation}
where $t$ is a reference point and
\begin{eqnarray}
Q^{2}(r)&=&R(r)-\frac{1}{4r^{2}}\nonumber\\
&=&\frac{r^{4(d-3)}}{\Delta^{2}}\left[E^{2}-V_{1}-\frac{1}{4r^{2}}+\frac{(d^{2}-5d+7)r_{h}^{d-3}}
{2r^{d-1}}-\frac{(d-2)^{2}r_{h}^{2(d-3)}}{4r^{2(d-2)}}\right] .
\end{eqnarray}
As $|E|\rightarrow\infty$, the zeros of $Q^{2}(r)$, or the turning points, $t_{n}$, in this WKB approximation, are close to the origin in the complex $r$-plane. In this limit,
\begin{equation}
Q^{2}(r)\sim\left(\frac{r^{d-3}}{r_{h}^{d-3}}\right)^{2}
\left[E^{2}-\frac{(d-2)^{2}r_{h}^{2(d-3)}}{4r^{2(d-2)}}\right] ,
\end{equation}
and the turning points are at:
\begin{equation}
Q^{2}(r)=0\Rightarrow
t^{2(d-2)}\sim\frac{(d-2)^{2}r_{h}^{2(d-3)}}{4E^{2}} .
\end{equation}
Asymptotically $E$ is very close to the negative imaginary axis, that is, $E\sim|E|e^{-i\pi/2}$, as such, the turning points can then be represented by:
\begin{equation}
t_{n}\sim\left[\frac{(d-2)^{2}r_{h}^{2(d-3)}}{4|E|^{2}}\right]^{1/2(d-2)}
e^{i(2n-1)\pi/2(d-2)} ,
\end{equation}
for $n=1,2,\cdots,2(d-2)$. Note that the equation giving the asymptotic quasinormal frequency involves two quantities, the first being the line integral from one turning point to the other. Here, we have:
\begin{eqnarray}
\gamma&\equiv&-\int_{t_{1}}^{t_{2}}d\xi\ \! Q(\xi)\nonumber\\
&=&-\int_{t_{1}}^{t_{2}}d\xi\left(\frac{\xi^{d-3}E}{r_{h}^{d-3}}\right)
\left[1-\frac{(d-2)^{2}r_{h}^{2(d-3)}}{4E^{2}\xi^{2(d-2)}}\right]^{1/2}\nonumber\\
&=&-\frac{1}{2}\int_{1}^{-1}\left(1-\frac{1}{y^{2}}\right)^{1/2}\nonumber\\
&=&-\frac{\pi}{2} ,
\end{eqnarray}
where we have made the change of variable $y=2E\xi^{(d-2)}/(d-2)r_{h}^{(d-3)}$. The other quantity is the closed contour integration around $r=r_{h}$:
\begin{equation}
\Gamma\equiv\oint_{r=r_{h}}d\xi\ \!Q(\xi)
=\oint_{r=r_{h}}d\xi\frac{r^{d-3}E}{r^{d-3}-r_{h}^{d-3}}
=\frac{2\pi i r_{h}E}{d-3} .
\end{equation}
The asymptotic quasinormal frequency is then given by the formula,
\begin{eqnarray}
e^{-2i\Gamma}=-(1+2\cos 2\gamma)&\Rightarrow&e^{4\pi
r_{h}E/(d-3)}=1\nonumber\\
&\Rightarrow&E_{n}=-i\left(\frac{d-3}{2r_{h}}\right)n ,
\end{eqnarray}
for large $n$. In terms of the corresponding Hawking temperature, $T_{H}=(d-3)/4\pi r_{h}$,
\begin{equation}
E_{n}=-i2\pi T_{H}n.
\end{equation}
We therefore obtain a vanishing real part for the frequency and find that the spacing of the imaginary part goes to $2\pi T_{H}$ regardless of the dimension. Note that these results are in accordance with that of integral spin fields in higher-dimensional Schwarzschild spacetimes \cite{Motl,SDas,Natario}.


\section{Concluding remarks}

\par In this paper we have presented new results for the QNMs of a massless Dirac field on a bulk $d$-dimensional background, see Figure \ref{QNModd} and \ref{QNMeven}.
The results can be compared with the brane-localized results of
reference \cite{BraneQNMs}, revealing that bulk fermion modes
result in much larger damping rates (in both cases larger $d$
results in greater damping rates). Some words of caution are
necessary, as can be seen for example in the $d=10$ result, which
plots the $l=0,1,2$ and $3$ channels. In the next angular momentum
channel, $l=4$, the point $n=l$, crosses the imaginary axis. The
presence of a branch cut along this axis would force us to choose
the positive imaginary value in accordance with the constraint
equation (\ref{eqn:alpha}). However this does not signify that
there are modes which are unstable, but indicates, as can be
deduced from our asymptotic large $n$ analysis, that the WKB
approximation is breaking down.
\par Furthermore, if a more detailed analysis such as along the lines of Leaver's approach \cite{Leaver} confirms the result that some of the QNMs have real part equal to zero
(algebraically special frequencies), as our WKB results imply for
$n\gtrsim l $, then such modes would be
 {\it transparent}.  However, as mentioned, we cannot say too much based upon the WKB approximation in this vicinity.

\par Also, given that the BH damping rate increases with dimension
$d$, see Figure \ref{QNMwithdim}, we can naively infer that the
amount of energy available to be radiated as Hawking radiation
will increase with $d$. These issues are left for further study
\cite{Forth}.

\par In section IV we investigated the large angular momentum limit and found an asymptotic result that agrees with the $d=4$ result.
In section V we discussed the remaining issue of asymptotic
quasinormal frequencies. The expected result in $d$-dimensions is
obtained in such a limit. Note that recently Dirac QNMs have been
analyzed under the influence of Lorentz violation \cite{Chen}. It
is of interest that the authors introduce a $\gamma_5$ term in
their equations similar to the one found in our reduced
2-dimensional Dirac equation (\ref{eqn:2Ddirac}).

\par Finally, it may be worth mentioning that split fermion theories often have massive fermions in the bulk, examples of this are the higher order modes of the quarks and leptons, which are freed from the localizing potential at high energy. Other examples are the bulk Higginos and gauginos in SUSY formulations. In addition to our results, depending on the energy scale involved, investigations of these massive Dirac QNMs would also be of interest.

\section*{Acknowledgments}

\par HTC was supported in part by the National Science Council of the Republic of China under the Grant NSC 95-2112-M-032-013. JD wishes to thank Dr. G. C. Joshi for his advice and supervision during the production of this work. The authors would also like to thank Prof. Misao Sasaki for the guidance and advice given during the many fruitful discussions we had with him.

\end{document}